\documentclass[%
reprint,
superscriptaddress,
 amsmath,amssymb,
 aps,
]{revtex4-2}

\usepackage[utf8]{inputenc}
\usepackage{graphicx}
\usepackage{dcolumn}
\usepackage{bm}
\usepackage{hyperref}
\hypersetup{
    colorlinks=true,
   allcolors=blue,
    }
\usepackage[dvipsnames]{xcolor}

\usepackage{xcolor}
\usepackage{physics}
\graphicspath{{./figs/}}

\begin{document}
\title{Critical quantum sensing via backaction-evading continuous measurement}

\author{Cheng Zhang}
\affiliation{Research Center for Quantum Physics and Technologies, Inner Mongolia University, Hohhot 010021, China}
\affiliation{School of Physical Science and Technology, Inner Mongolia University, Hohhot 010021, China}

\author{Mauro Cirio}
\affiliation{Graduate School of China Academy of Engineering Physics, Haidian District, Beijing, 100193, China}

\author{Xin-Qi Li}
\affiliation{Research Center for Quantum Physics and Technologies, Inner Mongolia University, Hohhot 010021, China}
\affiliation{School of Physical Science and Technology, Inner Mongolia University, Hohhot 010021, China}

\author{Pengfei Liang}
\email{pfliang@imu.edu.cn}
\affiliation{Research Center for Quantum Physics and Technologies, Inner Mongolia University, Hohhot 010021, China}
\affiliation{School of Physical Science and Technology, Inner Mongolia University, Hohhot 010021, China}

\date{\today}
\begin{abstract}
Continuous monitoring of open quantum-optical systems offers a promising route towards quantum-enhanced sensing. In such sensing schemes, the ultimate precision limit is dictated by the quantum Cram\'er-Rao bound. However, reaching this fundamental bound with established continuous measurement techniques has remained an outstanding challenge. Here, we present an approach tailored to open quantum-optical sensors operating near dissipative critical points. The key insight of our approach is to exploit a synergy between the intrinsic critical dynamics and the unavoidable quantum measurement backaction of continuous homodyne detection. By carefully aligning the homodyne phase to the squeezing angle of the sensor state, quantum backaction can be evaded in the critical dynamics, thereby significantly boosting the attainable precision. We demonstrate the broad applicability of this approach across several sensor models and for both single- and multi-parameter estimation, showing that it consistently narrows the gap toward the ultimate precision bound.
\end{abstract}

\pacs{}
\maketitle

\section{Introduction}
A fundamental principle of quantum mechanics is that quantum measurement inevitably disturbs the system being observed~\cite{Nielsen_Chuang_2010}. Beyond its conceptual importance, controlling and harnessing such measurement backaction is crucial for a wide range of quantum technologies~\cite{PhysRevLett.86.5188,PhysRevLett.91.250801,PhysRevA.70.052102,PhysRevA.70.052324,PhysRevA.69.032109,Walther2005,PhysRevA.71.032318,Benjamin2009,PhysRevB.79.035315,Blok2014,PhysRevLett.126.090503,MeasurementBasedQuantumComputation}. In quantum metrology, the quantum backaction of continuous measurements~\cite{Wiseman_Milburn_2009,Serafini,ALBARELLI2024129260} has attracted considerable interest, proving instrumental in tasks such as generating stable spin-squeezing~\cite{PhysRevLett.91.250801,PhysRevA.70.052102,PhysRevA.70.052324,PhysRevA.69.032109,PhysRevA.79.062107,Albarelli_2017} and mitigating detrimental local dephasing noises~\cite{Albarelli2018restoringheisenberg,PhysRevLett.125.200505}. 

Driven quantum-optical setups~\cite{Baumann2010,Castanos_2012,RevModPhys.85.553,PhysRevLett.113.020408,PhysRevA.90.022111,PhysRevLett.118.247402,PhysRevA.97.013825} constantly emit radiation quanta into their environment, which can be continuously monitored via well-established techniques in quantum optics~\cite{Wiseman_Milburn_2009,Serafini,ALBARELLI2024129260}, e.g., photon counting, homodyne, and heterodyne detection. As a result, they serve as ideal platforms for implementing sensing protocols based on continuous measurement~\cite{PhysRevA.87.032115,PhysRevA.89.052110,PhysRevA.102.063716,PhysRevA.103.032406}. The ultimate precision achievable by such protocols is dictated, through the quantum Cram\'er-Rao bound, by the global quantum Fisher information (QFI)~\cite{PhysRevLett.112.170401,PRXQuantum.3.010354,PhysRevX.13.031012} which quantifies the distinguishability of the underlying joint system-environment state. However, temporal correlations between the radiation quanta generally prevent saturation of this bound 
when using conventional time-local continuous measurements. To address this challenge, Yang {\it et al.} recently proposed a cascaded architecture capable of converting the correlated output field of the sensor into an uncorrelated field at the decoder~\cite{PhysRevX.13.031012}. Despite this important progress, there remains a pressing need for alternative approaches that can improve precision towards the quantum Cram\'er-Rao bound in open quantum sensors under realistic continuous measurement conditions. 

In this work, we show that by evading quantum measurement backaction, one can narrow this gap in open quantum-optical sensors featuring dissipative critical points. Our approach, consisting in continuously monitoring the sensor emission field via  homodyne detection~\cite{Wiseman_Milburn_2009,Serafini,ALBARELLI2024129260} and optimizing the homodyne phase, capitalizes on a crucial backaction-evading property of the sensor dynamics. This property  emerges from the interplay between measurement backaction and dissipative criticality, giving rise to special critical points around which the classical Fisher information (CFI) is significantly enhanced. We demonstrate 
the general applicability of the approach  
with three examples: frequency estimation with the open Kerr parametric oscillator (KPO)~\cite{PhysRevLett.36.1135,Drummond_1980,PhysRevA.94.033841,PhysRevA.95.012128,PhysRevA.98.042118,PhysRevA.106.033707}, magnetometry with the open Lipkin-Meshkov-Glick (LMG)~\cite{PhysRevLett.99.050402,PhysRevA.103.013306,PhysRevB.107.104201} model, and multi-frequency estimation with the open Dicke model~\cite{PhysRevLett.118.123602,PhysRevLett.120.183603,Kirton2019}.

\begin{figure}%[t!]
\includegraphics[clip,width=8.5cm]{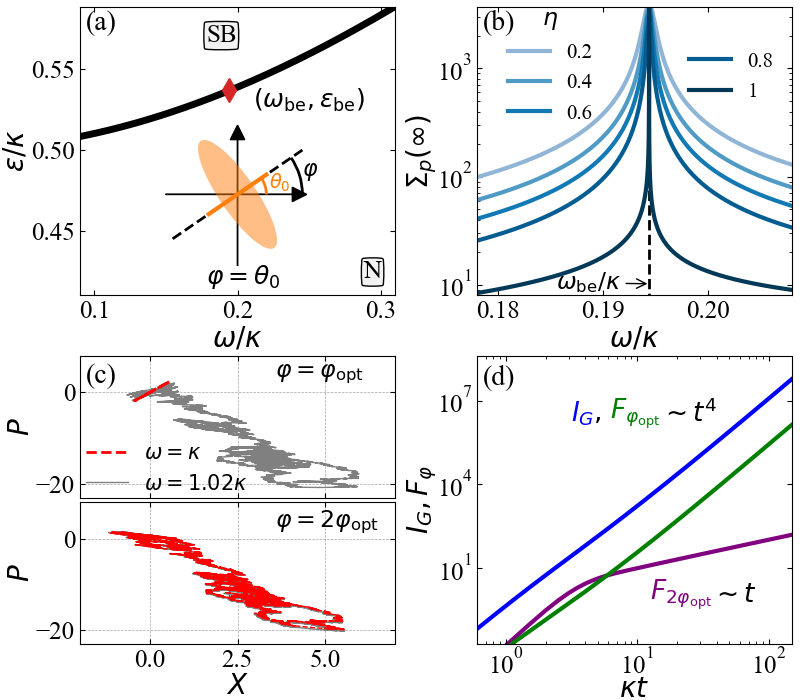}
\caption{(a) Phase diagram of the open KPO. The black solid curve marks the phase boundary separating the normal phase (N) and the symmetry-broken phase (SB). The inset illustrates the phase-matching condition~(\ref{eq:phasecondition}) for the emergence of backaction-evading points at $(\omega_\text{be},\epsilon_\text{be})$ [given by Eq.~(\ref{eq:epsilonomega})], where $\theta_0$ denotes the squeezing angle of the unconditional steady-state in the normal phase and $\varphi$ is the homodyne phase (here $\varphi=0.6$). (b) Stationary covariance $\Sigma_p(\infty)$ along the phase boundary $\epsilon=\epsilon_c(\omega)$ for $\varphi=0.6$. The other covariance components $\Sigma_x(\infty)$ and $\Sigma_{xp}(\infty)$ show similar behavior, see~\cite{suppmat}. (c) Trajectories of $\mathbf{r}$ in phase space. The red curves correspond to $\varphi=\varphi_\text{opt}$ (upper panel) and $\varphi\neq\varphi_\text{opt}$ (lower panel), both at frequency $\omega=\kappa$ and driving strength $\epsilon_c(\kappa)$. The gray curves are obtained at a slightly varied frequency $\omega=1.02\kappa$, with driving strength $\epsilon_c(1.02\kappa)$. (d) Long-time scaling of the CFI $F_\varphi$ and the global QFI $I_G$ for $\omega=\kappa$ and $\epsilon=\epsilon_c(\kappa)$. Here $\varphi_\text{opt}=\tan^{-1}(0.5)/2$ and the detection efficiency is $\eta=1$.
}\label{fig:fig1}
\end{figure}

\section{Frequency estimation with the open KPO}
As a first illustrative example, we consider estimating the frequency $\omega$ of the open KPO, focusing on the regime of vanishing Kerr nonlinearity. The oscillator dynamics is described by the Lindblad master equation~\cite{PhysRevLett.36.1135,Drummond_1980,PhysRevA.94.033841,PhysRevA.95.012128,PhysRevA.98.042118,PhysRevA.106.033707}
\begin{equation}\label{eq:me}
\partial_t\rho= \mathcal{L}\rho = -i[H_\omega,\rho] + \kappa \mathcal{D}[a]\rho, 
\end{equation}
where the Hamiltonian $H_\omega = \omega a^\dagger a + \epsilon(a^{\dagger2} + a^2)/2$ generates the unitary evolution, in terms of the two-photon driving amplitude  $\epsilon>0$, and where the dissipator
$\mathcal{D}[a]\rho = a\rho a^\dagger - (a^\dagger a\rho + \rho a^\dagger a)/2$ accounts for photon loss at the rate $\kappa$. 

The open KPO exhibits two different kinds of dissipative phase transitions, depending on the sign of $\omega$~\cite{PhysRevLett.36.1135,Drummond_1980,PhysRevA.94.033841,PhysRevA.95.012128,PhysRevA.98.042118,PhysRevA.106.033707}. Specifically, for $\omega>0$ ($\omega<0$), the system undergoes a continuous (first-order) phase transition from the normal phase to the $\mathbb{Z}_2$ symmetry-broken phase. The phase boundary for the continuous phase transition is given by  $\epsilon=\epsilon_c(\omega)$, where $\epsilon_c(\omega) = \sqrt{\omega^2+\kappa^2/4}$~\cite{DiCandia2023,PhysRevLett.133.040801}, as shown in Fig.~\ref{fig:fig1}(a).  Since quantum criticality is central to this work, we focus on the positive-frequency regime $\omega>0$.

Our approach employs continuous homodyne detection of an oscillator quadrature $x_\varphi=x\cos\varphi+p\sin\varphi$, where $\varphi$ is the homodyne phase, and $x=(a+a^\dagger)/\sqrt2$, $p=i(a^\dagger -a)/\sqrt2$ are the standard orthogonal quadratures. Under this monitoring, the oscillator state $\rho_c$ becomes conditioned on the measurement record and evolves according to the stochastic master equation (SME)~\cite{Wiseman_Milburn_2009,Serafini,ALBARELLI2024129260}
\begin{equation}\label{eq:sme}
d\rho_c = \mathcal{L}\rho_c\,dt + \sqrt{\eta\kappa}\,\mathcal{H}[ae^{-i\varphi}]\rho_c\,dw,
\end{equation}
where $\mathcal{H}[ae^{-i\varphi}]\rho_c = ae^{-i\varphi}\rho_c + \rho_ca^\dagger e^{i\varphi} - \langle ae^{-i\varphi} + a^\dagger e^{i\varphi}\rangle_c\rho_c$ represents the measurement backaction and where $\eta\in[0,1]$ is the detection efficiency. The notation $\langle o\rangle_c = \text{Tr}[o\rho_c]$ denotes the conditional expectation of an operator $o$. The Wiener increment $dw$ satisfies $\mathbb{E}[dw]=0$ and $\mathbb{E}[(dw)^2]=dt$, and the recorded photocurrent reads $dy_t = \sqrt{2\kappa\eta}\langle x_\varphi\rangle_cdt + dw$.

In the normal phase ($\epsilon<\epsilon_c$), the conditional state retains its Gaussian character for any Gaussian initial state. We denote by $\Sigma_o= 2(\langle o^2\rangle_c - \langle o\rangle_c^2)$ the variance of the operator $o$, and by $\Sigma_{o_1o_2}= \langle\{o_1,o_2\}\rangle_c - 2\langle o_1\rangle_c\langle o_2\rangle_c$ the covariance between the operators $o_1$ and $o_2$. The conditional state is then fully characterized by its mean vector $\mathbf{r} = (X,P)^\intercal$, with $X=\langle x\rangle_c$ and $P=\langle p\rangle_c$, and its covariance matrix $\mathbf\Sigma = (\Sigma_x, \Sigma_{xp}; \Sigma_{xp}, \Sigma_p)$, which obey the differential equations~\cite{Serafini,PhysRevA.95.012116}
\begin{subequations}\label{eq:rsigmaeqs}
\begin{align}
\displaystyle d\mathbf{r} &\displaystyle= \mathbf{A} \mathbf{r}dt + \sqrt{\frac{\eta\kappa}{2}}(\mathbf\Sigma-\mathbf{I})\mathbf{v}dw, \label{eq:rsigmaeqs1} \\
\displaystyle\frac{d\mathbf\Sigma}{dt} &\displaystyle= \mathbf{A}\mathbf\Sigma + \mathbf\Sigma \mathbf{A}^\intercal + \mathbf{D} - \eta\kappa(\mathbf\Sigma - \mathbf{I})\mathbf{B}(\mathbf\Sigma - \mathbf{I}), \label{eq:rsigmaeqs2}
\end{align}
\end{subequations}
where $\mathbf{A} = (-\kappa/2, \omega-\epsilon; -\omega-\epsilon, -\kappa/2)$ is the drift matrix, $\mathbf{D}=\text{diag}(\kappa,\kappa)$ the diffusion matrix, $\mathbf{v}=(\cos\varphi,\sin\varphi)^\intercal$ and $\mathbf{B}=\mathbf{v}\mathbf{v}^\intercal$. A detailed derivation of Eq.~(\ref{eq:rsigmaeqs}) is provided in the Supplemental Material~\cite{suppmat}. We restrict the discussion to the normal phase and the phase boundary.

\section{Backaction-evading property}
The core of the approach lies in further optimizing the homodyne phase $\varphi$, a step that exploits the following crucial property of the conditional dynamics described by Eq.~(\ref{eq:rsigmaeqs}): For any homodyne phase $\varphi\in(0,\pi/4)$, in the long-time limit $t\to\infty$, the covariance $\mathbf{\Sigma}(t)$ remains bounded and saturates to finite stationary values along the critical phase boundary $(\omega,\epsilon_c(\omega))$, with one remarkable exception at the ``backaction-evading" point
\begin{equation}\label{eq:epsilonomega}
(\omega_\text{be},~\epsilon_\text{be}) = \left( \frac{\kappa}{2}\cot(2\varphi),~\frac{\kappa}{2\sin(2\varphi)} \right), 
\end{equation}
where $\mathbf{\Sigma}(t)$ diverges in time, irrespective of the efficiency $\eta$. This property is illustrated in Fig.~\ref{fig:fig1}(b) and rigorously proven in~\cite{suppmat}.

This feature can be justified with the following physically motivated argument. In Markovian open systems, a dissipative phase transition is often characterized by a closing Liouvillian gap, i.e., by the vanishing of the smallest damping rate of the Liouvillian $\mathcal{L}$. This corresponds to an increasingly slow relaxation towards equilibrium which, for the open KPO, can lead to the divergence of a quadrature variance. 
For example, in the normal phase the steady state $\rho_\text{ss}$ of Eq.~(\ref{eq:me}) is a squeezed state with squeezing angle $\theta_0 = \tan^{-1}(\kappa/2\omega)/2$. In this case, the squeezed quadrature $x_{\theta_0} = x\cos\theta_0 + p\sin\theta_0$ has variance $\Sigma_{x_{\theta_0}}=\epsilon_c/(\epsilon_c+\epsilon)$ below the coherent-state limit of $1$, while the anti-squeezed conjugate $p_{\theta_0} = -x\sin\theta_0 + p\cos\theta_0$ has variance $\Sigma_{p_{\theta_0}}=\epsilon_c/(\epsilon_c-\epsilon)$ above $1$, and it diverges as $\epsilon\nearrow\epsilon_c$.

Quantum backaction counteracts this behavior. In fact, measuring $x_\varphi$ tends to localize the conditional state in this quadrature, thereby reducing its variance~\footnote{The backaction contribution to the covariance, $-\eta\kappa(\mathbf\Sigma - \mathbf{I})\mathbf{B}(\mathbf\Sigma - \mathbf{I})$ from  Eq.~(\ref{eq:rsigmaeqs2}), is negative semi-definite: For any vector $\mathbf{u}\in\mathbb{R}^2$, the inequality $-\eta\kappa \mathbf{u}^\intercal(\mathbf\Sigma - \mathbf{I})\mathbf{B}(\mathbf\Sigma - \mathbf{I})\mathbf{u} = -\eta\kappa \bigl[\mathbf{v}^\intercal(\mathbf\Sigma - \mathbf{I})\mathbf{u}\bigr]^2 \le0$ holds}. This explains why critical fluctuations become bounded in the numerical results of Fig.~\ref{fig:fig1}(b). However, a notable exception arises when the homodyne phase $\varphi$ aligns with the squeezing angle $\theta_0$ for the steady state $\rho_\text{ss}$, i.e., 
\begin{equation}\label{eq:phasecondition}
\varphi = \theta_0.
\end{equation}
Substituting $\theta_0 = \tan^{-1}(\kappa/2\omega)/2$ into this phase-matching condition directly yields the backaction-evading condition in Eq.~(\ref{eq:epsilonomega}). This equivalence shows that the unbounded fluctuation at the points in Eq.~(\ref{eq:epsilonomega}) is a direct consequence of the well-known backaction‑evading mechanism in quantum physics~\cite{Scully} which, here, operates in the context of a dissipative phase transition, thereby also justifying our nomenclature choice. 
In~\cite{suppmat}, we complement this qualitative interpretation with a perturbative analysis for $\eta\ll1$ and further present an experimental scheme for detecting backaction-evading points via the integrated current $y_t=\int_0^tdy_t$.

\begin{figure}%[t!]
\includegraphics[clip,width=8.5cm]{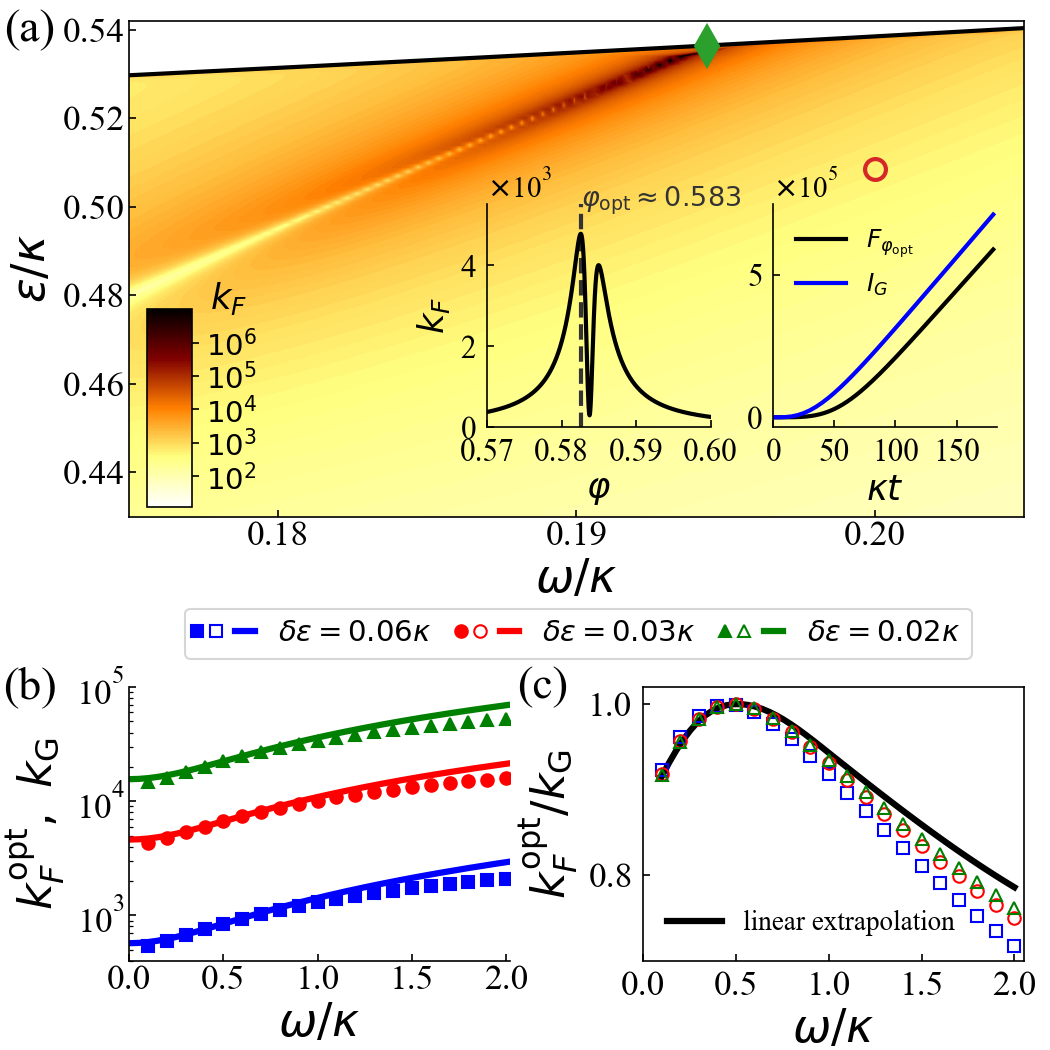}
\caption{(a) The long-time growth rate $k_F$ of the CFI $F_{\varphi}$ as a function of $\omega$ and $\epsilon$. The green diamond corresponds to the backaction-evading point for $\varphi=0.6$. Left inset: Optimization of $k_F$ over the phase $\varphi$ for $(\omega,\delta\epsilon)=(0.2\kappa,0.03\kappa)$ [red  circle]. The optimal phase is $\varphi_\text{opt}\approx0.583$ (gray dashed line). Right inset: Time evolution of the CFI $F_{\varphi_\text{opt}}$ and the corresponding global QFI $I_G$. (b) The growth rates $k_F^\text{opt}$ (symbols) and $k_G$ (lines) and (c) their ratio $k_F^\text{opt}/k_G$ (empty symbols) as functions of $\omega$ for $\delta\epsilon/\kappa = 0.06$, $0.03$ and $0.02$. The solid curve in (c) represents the linear extrapolation of $k_F^\text{opt}/k_G$ to $\delta\epsilon=0$. All simulations assume ideal detection efficiency $\eta=1$. 
}\label{fig:CFIprofile}
\end{figure}

\section{Sensing performance}
We now demonstrate that the above backaction-evading property allows us to optimize the phase $\varphi$, significantly enhancing the associated CFI,  $F_\varphi$~\cite{Albarelli2018restoringheisenberg,PhysRevA.87.032115}, and further narrowing the gap towards the global QFI,  $I_G$, of the joint oscillator-environment state \cite{PhysRevLett.112.170401,PRXQuantum.3.010354,PhysRevX.13.031012}. More specifically, the variance of any unbiased estimator $\hat\omega$ obeys the bound
\begin{equation}\operatorname{Var}[\hat\omega] \ge 1/F_\varphi \ge 1/ I_G.
\end{equation}
Here, since homodyne detection is a Gaussian measurement~\footnote{The probability density of the homodyne photocurrent $dy_t$ is Gaussian.}, the CFI can be evaluated via~\cite{PhysRevA.95.012116}
\begin{equation}\label{eq:Fhd}
F_\varphi = 2\eta\kappa\int_0^t d\tau\,\mathbb{E}\bigl[(\partial_\omega\mathbf{r}^\intercal)\mathbf{B}(\partial_\omega\mathbf{r})\bigr]. 
\end{equation}
In~\cite{suppmat}, we derive a set of deterministic differential equations to calculate the integrand, thereby avoiding the need for computationally expensive ensemble averages.
 
The global QFI is given by~\cite{PhysRevLett.112.170401,PRXQuantum.3.010354,PhysRevX.13.031012}
\begin{equation}\label{eq:IGdef}
I_G = 4\partial_{\omega_1}\partial_{\omega_2}\ln\,\mathcal{F}(\omega_1,\omega_2)\vert_{\omega_1=\omega_2=\omega},
\end{equation}
where $\mathcal{F}(\omega_1,\omega_2)$ is the quantum fidelity between the joint states for frequency values $\omega_1$ and $\omega_2$. While this can be computed by solving a generalized master equation~\cite{PhysRevLett.112.170401,PRXQuantum.3.010354,PhysRevX.13.031012}, in \cite{suppmat} we derive the equivalent integral representation
\begin{equation}\label{eq:intgrep}
I_G = 4\text{Im}\int_0^tds\,\operatorname{Tr}[a^\dagger a\partial_{\delta\omega}\,\mu] - 4\left(\int_0^td\tau\,\operatorname{Tr}[a^\dagger a\rho]\right)^2, 
\end{equation}
where $\mu$ is a generalized density matrix defined in~\cite{suppmat}, and $\partial_{\delta\omega}\,\mu$ denotes its derivative with respect to $\delta\omega = (\omega_2-\omega_1)/2$. These results allow us to efficiently calculate both quantities. In what follows, we focus on ideal homodyne measurement ($\eta=1$).

At criticality ($\delta\epsilon=0$), the optimal phase for estimating $\omega$ is given by $\varphi=\varphi_\text{opt}$ such that $\omega=\omega_\text{be}(\varphi_\text{opt})$, where $\varphi_\text{opt}=\tan^{-1}(\kappa/2\omega)/2$ is obtained by inverting the first expression in Eq.~(\ref{eq:epsilonomega}). At $\varphi=\varphi_\text{opt}$, the trajectory of $\mathbf{r}$ in phase space becomes highly sensitive to variations in $\omega$; whereas for $\varphi\neq\varphi_\text{opt}$ it is not, as illustrated in Fig.~\ref{fig:fig1}(c). Since  $F_\varphi$ is related to $\partial_\omega\mathbf{r}$ [see Eq.~(\ref{eq:Fhd})], this sensitivity results in the long-time scaling $F_{\varphi_\text{opt}}\sim t^4$, which coincides with that of the QFI, $I_G\sim t^4$ (see the proof in~\cite{suppmat}). For all other values of $\varphi$, the scaling is less favorable, $F_\varphi\sim t$, as exemplified in Fig.~\ref{fig:fig1}(d) for the representative case $\varphi=2\varphi_\text{opt}$.

In the off-critical case ($\delta\epsilon>0$), both $F_\varphi$ and $I_G$ exhibit linear temporal scaling 
$F_\varphi\sim k_Ft$ and $I_G\sim k_G t$, whose growth rates are given by 
$k_F = 2\eta\kappa\lim_{t\to\infty} \mathbb{E}\left[(\partial_\omega\mathbf{r}^\intercal)\mathbf{B}(\partial_\omega\mathbf{r})\right]$ and (see~\cite{suppmat} for detailed derivations and numerical validation) 
\begin{equation}\label{eq:kGsol}
k_G = 2\epsilon^2\frac{(\epsilon_c^2-2\kappa\omega^2+2\omega^2)(\epsilon_c^2-\epsilon^2)+\kappa^2\epsilon_c^2}{\kappa(\epsilon_c^2-\epsilon^2)^3}. 
\end{equation}
In Fig.~\ref{fig:CFIprofile}(a), we plot the landscape of $k_F$ for a fixed $\varphi=0.6$. 
A notable feature is the presence of a region of enhanced $k_F$, which ends at the corresponding backaction-evading point (green diamond) on the phase boundary (black solid line). Thanks to its dependence on $\varphi$, the location of this region can be tuned to  include the coordinate $(\omega,\epsilon)$  determined by the frequency $\omega$ to be estimated, and the control parameter $\epsilon$. In this way, it is possible to choose an optimal value for $\varphi$ which optimizes the estimation of $\omega$ for a given $\epsilon$. The resulting optimal phase is then defined as $\varphi_\text{opt}=\operatorname{argmax}_{\varphi}k_F$, and the corresponding rate is denoted as $k_F^\text{opt}$. This optimization is illustrated in the left inset of  Fig.~\ref{fig:CFIprofile}(b) for $(\omega,\delta\epsilon)=(0.2\kappa,0.03\kappa)$ [red circle]. The optimal value is found at $\varphi_\text{opt}\approx0.583$. The right inset compares the CFI evaluated with this optimal phase to the corresponding global QFI. Remarkably, their long-time growth rates nearly coincide, confirming the effectiveness of our approach.

To further assess the sensing performance,  we plot in Fig.~\ref{fig:CFIprofile}(b)  $k_F^\text{opt}$ and $k_G$, and in Fig.~\ref{fig:CFIprofile}(c) their ratio $k_F^\text{opt}/k_G$, all as functions of $\omega$. The best performance occurs at $\omega = 0.5\kappa$ where $k_F^\text{opt}/k_G = 1$, indicating that the conditional state $\rho_c$ becomes as informative as the full oscillator-environment state in the long-time limit, since both yield the same amount of information per unit time. Over a broad range around this sweet spot, $k_F^\text{opt}$ remains strongly enhanced. Approaching the phase boundary $\epsilon\nearrow\epsilon_c$, the ratio $k_F^\text{opt}/k_G$ converges
to the solid line in Fig.~\ref{fig:CFIprofile}(b),  obtained by extrapolating $k_F^\text{opt}/k_G$ to $\delta\epsilon=0$.

In~\cite{suppmat}, we present additional results for this setup, including an experimental feasibility analysis, an extension to  continuous general-dyne measurements~\cite{Serafini,PhysRevA.95.012116}, applications for non-ideal detectors, and a performance comparison to photon counting. In the following, we present two further  applications, one in magnetometry and one in multi-frequency estimation.

\section{Magnetometry with the open LMG model}
Here, we consider a magnetometry setup based on the open LMG model~\cite{PhysRevLett.99.050402,PhysRevA.103.013306,PhysRevB.107.104201}. The model describes a collection of coupled $N$ spin-half with Hamiltonian
$H^N_\text{LMG} = h S_z - (2J/N)S_x^2$, where $S_{\alpha} = \sum_{j=1}^N \sigma_{j,\alpha}/2$ for $\alpha=x,y,z$, with $\sigma_{j,\alpha}$ the Pauli operators for the $j$th spin. Here, $J$ is the coupling strength, and $h$ is the Zeeman field to be estimated. 
We model the effects of the environment by replacing the damping term in  Eq.~(\ref{eq:me}) with the collective dissipator $(\kappa/N)\mathcal{D}[S_-]\rho$, where $S_- = \sum_{j=1}^N \sigma_{j,-}$ with $\sigma_{j,\pm} = (\sigma_{j,x}\pm i\sigma_{j,y})/2$. Accordingly, the backaction term in the SME~(\ref{eq:sme}) should be changed to $\sqrt{\eta\kappa/N}\mathcal{H}[S_-e^{-i\varphi}]\rho_c$. 

Applying the Holstein-Primakoff  transformation~\cite{PhysRev.58.1098} $S_+ = \sqrt{N-a^\dagger a}a^\dagger$, $S_z = a^\dagger a - N/2$ and taking the thermodynamic limit $N\to\infty$, the LMG Hamiltonian reduces to $H_\text{LMG}^{\infty} = (h-J)a^\dagger a - (J/2)(a^2+a^{\dagger2})$;  the damping term in Eq.~(\ref{eq:me}) reduces to $\kappa\mathcal{D}[a]\rho$, and the backaction term in Eq.~(\ref{eq:sme}) to $\sqrt{\eta\kappa}\mathcal{H}[ae^{-i\varphi}]\rho_c$. As a result, the LMG model in the thermodynamic limit is formally equivalent to the KPO model, with the equivalence established explicitly via the mapping $\omega \to h-J$, $\epsilon \to -J$. All results presented above for estimating $\omega$ therefore apply equally to the estimation of $h$ in the LMG magnetometer. 

\begin{figure}%[t!]
\includegraphics[clip,width=8.5cm]{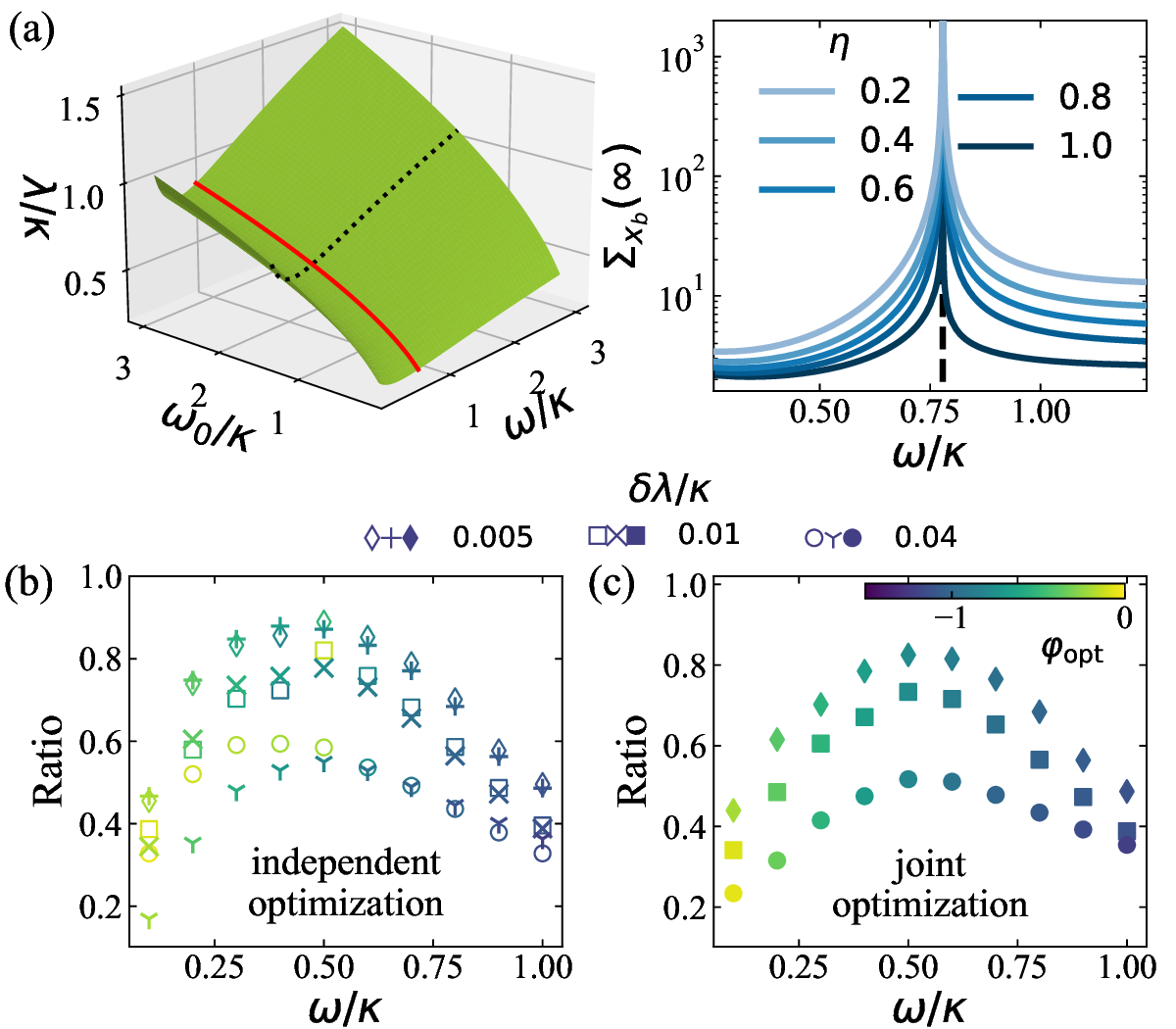}
\caption{(a) Left panel: Phase diagram of the open Dicke model. The colored surface is the phase boundary, below (above) which lies the normal (superradiant) phase. The red solid curve shows the backaction-evading line for $\varphi=-1$. Right panel: Stationary covariance $\Sigma_{x_b}(\infty)$ along the dotted linecut in the left panel. The backaction-evading point is marked by the dashed line. (b) The ratios $k_{F,1}^\text{opt}/k_{G,1}$ (empty symbols) and $k_{F,2}^\text{opt}/k_{G,2}$ (Plus, cross, and Y symbols), and (c) $k_{F,3}^\text{opt}/k_{G,3}$ (filled symbols) for different boundary proximities $\delta\lambda$ at fixed $\omega_0=0.5\kappa$. The colors show the optimal homodyne phase $\varphi_\text{opt}$. All simulations assume ideal detection efficiency $\eta=1$. 
}\label{fig:dicke}
\end{figure}

\section{Multi-frequency estimation with the open Dicke model}
As a final example, we consider multi-frequency estimation using the open Dicke model~\cite{PhysRevLett.118.123602,PhysRevLett.120.183603,Kirton2019}, again described by the master equation~(\ref{eq:me}) but now with the Hamiltonian $H^N_\text{Dicke} = \omega a^\dagger a + \omega_0 S_z + 2\lambda/\sqrt{N}(a+a^\dagger)S_x$, where $\omega_0$ denotes the atomic frequency and $\lambda$ the light-matter coupling strength. Applying the Holstein-Primakoff transformation $S_+ = \sqrt{N-b^\dagger b}b^\dagger$, $S_z = b^\dagger b - N/2$ and taking the limit $N\to\infty$, the Hamiltonian reduces to a two-mode form $H_\text{Dicke}^\infty = \omega a^\dagger a + \omega_0 b^\dagger b + \lambda(a+a^\dagger)(b+b^\dagger)$. The model undergoes a dissipative phase transition from the normal phase to the superradiant phase at the critical coupling $\lambda_c = \sqrt{(\kappa^2+4\omega^2)\omega_0/\omega}/4$~\cite{suppmat}. This example demonstrates the applicability of our approach to multi-mode sensor models and multi-parameter estimation. 

The cavity emission is monitored continuously via homodyne detection, so that the conditional dynamics still obeys the SME~(\ref{eq:sme}) with backaction-evading condition given by~\cite{suppmat} 
\begin{equation}\label{eq:dickebecp}
(\omega_\text{be}, \lambda_\text{be}) = \left(-\frac{\kappa}{2}\tan\varphi, \frac12 \sqrt{\frac{-\kappa\omega_0}{\sin(2\varphi)}}\right). 
\end{equation}
Interestingly, for a given $\varphi$ this condition defines a backaction-evading line as demonstrated, and numerically verified, in Fig.~\ref{fig:dicke}(a).  

We consider the simultaneous estimation of $\omega$ and $\omega_0$. The mean-squared error matrix $\mathbf{\Delta}_{\boldsymbol{\hat\omega}} = \mathbb{E}[(\boldsymbol{\hat\omega} - \mathbb{E}[\boldsymbol{\hat\omega}])(\boldsymbol{\hat\omega} - \mathbb{E}[\boldsymbol{\hat\omega}])^\intercal]$ of an unbiased vector-valued estimator $\boldsymbol{\hat\omega} = (\hat\omega_0, \hat\omega)^\intercal$ satisfies the bound $\mathbf\Delta_{\boldsymbol{\hat\omega}} \succeq \mathbf{F}^{-1}_\varphi \succeq \mathbf{I}_G^{-1}$~\cite{Szczykulska03072016,PhysRevLett.126.200501}, valid for any $\varphi$, where $\succeq$ denotes the Loewner order, $\mathbf{F}_\varphi$ is the CFI matrix, and $\mathbf{I}_G$ is the global QFI matrix. We restrict our discussion to the normal phase $\lambda < \lambda_c$, where the Gaussian description is valid, and present details for calculating $\mathbf{F}$ and $\mathbf{I}_G$ in~\cite{suppmat}.

When one of the frequencies is of particular interest (the other is treated as unknown), the variance of the corresponding estimator is bounded by a diagonal element of the inverse CFI matrix, e.g., $\operatorname{Var}(\hat\omega_0) \ge [\mathbf{F}^{-1}_\varphi]_{11}$ and $\operatorname{Var}(\hat\omega) \ge [\mathbf{F}^{-1}_\varphi]_{22}$. We therefore consider independent optimization of each bound: $\min_\varphi [\mathbf{F}^{-1}_\varphi]_{11}$ and $\min_\varphi [\mathbf{F}^{-1}_\varphi]_{22}$. In the long-time limit where the CFI grows linearly with time, these minimizations are equivalent to maximizing the asymptotic rates
\begin{equation}\label{eq:seprates}
k_{F,1} = \lim_{t\to\infty} \frac{1}{t[\mathbf{F}^{-1}_\varphi]_{11}}, \qquad
k_{F,2} = \lim_{t\to\infty} \frac{1}{t[\mathbf{F}^{-1}_\varphi]_{22}}.
\end{equation}
Note that the optimal phases for $\omega_0$ and $\omega$ are generally different.

When both frequencies are equally important, a joint optimization of the precision matrix is required. Here we adopt the criterion $\min_\varphi \operatorname{Tr}[\mathbf{F}^{-1}_\varphi]$~\cite{PhysRevLett.126.200501}, corresponding to the smallest average variance. In the long-time limit, this is equivalent to maximizing
\begin{equation}\label{eq:jointrate}
k_{F,3} = \lim_{t\to\infty} \frac{1}{t\operatorname{Tr}[\mathbf{F}^{-1}_\varphi]}.
\end{equation}

We characterize the sensing performance by the optimal ratios $k_{F,1}^\text{opt}/k_{G,1}$, $k_{F,2}^\text{opt}/k_{G,2}$ for independent optimization [Fig.~\ref{fig:dicke}(b)], and by $k_{F,3}^\text{opt}/k_{G,3}$ for joint optimization [Fig.~\ref{fig:dicke}(c)],  where the asymptotic rates $k_{G,i}$ ($i=1,2,3$) are defined using Eqs.~(\ref{eq:seprates}) and (\ref{eq:jointrate}) with $\mathbf{I}_G$ in place of $\mathbf{F}_\varphi$. Here, the atomic frequency is fixed at $\omega_0=0.5\kappa$. For a small but non-vanishing boundary proximity $\delta\lambda = \lambda_c - \lambda$, the best performance is achieved near $\omega\approx\omega_0$ for both optimization scenarios (see~\cite{suppmat} for numerical results at other atomic frequencies). Importantly, as the transition is approached ($\delta\lambda \to 0$), the performance improves, with all ratios increasing towards unity.

\section{Conclusions}
In summary, we have considered quantum sensors built upon open quantum-optical models operating near dissipative critical points. We showed that, at criticality, it is possible to avoid quantum backaction when operating continuous homodyne detection on the sensor. We analyzed the benefits of this specific interplay between the properties of the measurement device and the state of the sensor, which  are encoded in an enhancement of the classical Fisher information able to narrow the gap in the quantum Cram\'er-Rao bound. We further demonstrated the general applicability of our approach across single- and multi-mode sensor models and for both single- and multi-parameter estimation. These results pave the way towards a practical route to near-optimal precision in critical sensing under realistic continuous measurements. Promising directions for future work include exploring the scaling behavior of the classical Fisher information at backaction-evading points beyond the Gaussian regime~\cite{PRXQuantum.3.010354}, and incorporating advanced estimation techniques~\cite{Amoros-Binefa_2021,k7nk-lrwd,amorosbinefa2025trackingtimevaryingsignalsquantumenhanced,PhysRevA.103.032406,Duan2025,zhang2026criticalityenhancedglobalfrequencysensing} for different sensing tasks. \nocite{Zhou,PRXQuantum.5.020201,PRXQuantum.4.010333}

\section{Acknowledgements}
C.Z. acknowledges support from
CPSF (Grants No. 2025M1773419). M.C. acknowledges support from NSFC (Grant No. 12088101) and NSAF (Grant No. U2330401). 

\bibliography{refs}

\begin{thebibliography}{64}
\bibitem [{sup()}]{suppmat}%
  \BibitemOpen
  \href@noop {} {\bibinfo {title} {See {S}upplemental {M}aterial for a
  derivation of the differential equation~(\ref{eq:rsigmaeqs}), a rigorous
  proof of the backaction-evading property, additional numerical results for
  the covariance components, an experimental scheme for detecting
  backaction-evading critical points, detailed calculations of the {CFI} and
  {QFI} for the {KPO}, an experimental feasibility analysis, an extension to
  continuous general-dyne measurement, a performance comparison to photon
  counting, and additional analytical and numerical results for the open
  {D}icke model. it also includes {R}efs.~[62-64].}}\BibitemShut {Stop}%
\end{thebibliography}

\end{document}